\documentclass[prd,aps,twocolumn,reprint,floatfix,nofootinbib,superscriptaddress]{revtex4}
\pdfoutput=1
\usepackage{amssymb,graphicx}
\usepackage{amsmath}
\usepackage{multirow}
\usepackage{epsfig}
\usepackage[usenames]{color} 
\usepackage[export]{adjustbox}

\newcommand{\beqn}{\begin{eqnarray}}
\newcommand{\eeqn}{\end{eqnarray}}
\newcommand{\beq}{\begin{equation}}
\newcommand{\eeq}{\end{equation}}

\def\mphi{m_{\phi}}

\def\pt{\tilde{p}}
\def\rt{\tilde{\rho}}

\begin{document}

\title{Spontaneous growth of vector fields in gravity}
\author{Fethi M.\ Ramazano\u{g}lu}
\affiliation{Department of Physics, Ko\c{c} University, \\
Rumelifeneri Yolu, 34450 Sariyer, Istanbul, Turkey}

\begin{abstract}
We show that the spontaneous scalarization scenario in scalar-tensor theories is a specific case of
a more general phenomenon.
The key fact is that the instability causing the spontaneous growth in scalars is due to the
nonminimal coupling in the theory, and not related to the nature of the scalar.
Another field with the same form of coupling undergoes
spontaneous growth as well. We explicitly demonstrate this idea for vectors, naming it
``spontaneous vectorization'', and study spherically symmetric neutron stars in such a theory. 
We also comment on other tensor fields the idea can be applied, naming the general mechanism ``spontaneous tensorization''.
\end{abstract}
\maketitle
\section{Introduction}
Spontaneous scalarization is known as a phenomenon in certain scalar-tensor theories where the 
scalar field vacuum is unstable near neutron stars (NSs), meaning arbitrary perturbations from vacuum would
grow to a stable non-zero scalar field cloud~\cite{PhysRevLett.70.2220}.
These scalar fields die off away from the star, hence the theory passes the weak-field tests~\cite{Will:2006LR}.
More intriguingly, the modifications to general relativity (GR) near the star is order-of-unity, leading to
potentially large observable signatures in the strong-field regime~\cite{PhysRevLett.70.2220,Will:2006LR,2013PhRvD..87h1506B,2014PhRvD..89h4005S}.

The origin of spontaneous scalarization can be traced to a long wavelength tachyon instability
in the presence of matter as we will discuss in more detail~\cite{Ramazanoglu:2016kul}. The main point
of this paper is that this tachyon instability is due to the form of the nonminimal coupling term in the action of the theory, and
the scalar nature of the field is not important. This suggests that any field with a similar
coupling likely presents the same instability, and spontaneously grows. Hence, we name
this more general phenomenon \textit{spontaneous tensorization}.

Replacing the scalar field in the scalar-tensor theory with a vector field is a natural first step in generalizing spontaneous growth, and
we investigate such a theory in some detail. We indeed observe the vector field vacuum to be unstable when matter exists in the spacetime,
and study the properties of spontaneously vectorized NSs.
There are various alternative theories of gravity that involve vector fields~\cite{Horava:2009uw,Emparan2008,Jacobson:2000xp,
Moffat:2005si,Bekenstein:2004ne,Chamseddine:2013kea,Sebastiani:2016ras,BeltranJimenez:2013fca}.
There are also various scenarios in particle physics that predict existence of hitherto undetected vector or
pseudo-vector particles, especially those with ultralight masses that we are interested
~\cite{Arvanitaki:2009fg,Kodama:2011zc,Jaeckel:2010ni,Goodsell:2009xc}. The most relevant 
cases to our discussion in the literature are those with screening fields~\cite{BeltranJimenez:2013fca}.


Our main result is a new category of gravity theories that feature spontaneously growing tensor fields of different kinds with a common functional
form for their nonminimal matter coupling.
We emphasize that the study of a spontaneously growing vector field for which we spend considerable space below
acts as a demonstration of this idea, and is not the central theme itself.
We expect all these theories to share the appeal of spontaneous scalarization in terms of passing weak-field tests while providing 
order-of-unity deviations from GR in the strong-field. 
This aspect is of central importance in the age of gravitational wave science~\cite{PhysRevLett.116.061102} since they make contact
with observations considerably easier.
\section{The Origin of Spontaneous Scalarization}\label{sec_st}
Spontaneous scalarization is the prototypical example of spontaneous tensorization and we start with explaining
its physical mechanism. We follow \cite{Ramazanoglu:2016kul} where a more detailed discussion can be found.
The action for a scalar-tensor theory in the so called
Einstein frame is given by~\cite{PhysRevLett.70.2220}
\begin{align}\label{st_action}
 \frac{1}{16\pi} &\int d^4x \sqrt{-g} R -  \frac{1}{16\pi} \int d^4x \sqrt{-g} \left[ 2g^{\mu \nu} \partial_{\mu} \phi  \partial_{\nu} \phi
 + 2 m_{\phi}^2 \phi^2 \right] \nonumber \\
 &+ S_m \left[\psi_m, A^2(\phi) g_{\mu \nu} \right]
\end{align}
where $g_{\mu\nu}$ is the metric, $\phi$ is the scalar field and $m_{\phi}$ is 
the parameter coupling to the mass potential. $S_m$ is the matter action, where $\psi_m$
denote the matter degrees of freedom. The first line of the action is simply a massive scalar field living in GR, hence does
not constitute an \textit{alternative} theory of gravity. However, 
$\psi_m$ couple to a conformally scaled version of the metric, 
$\tilde{g}_{\mu\nu} = A^2(\phi) g_{\mu\nu}$, and this nonminimal coupling differentiates this theory from GR.
The scaled metric $\tilde{g}_{\mu\nu}$ defines the so-called Jordan frame, and is the metric physical observers
directly interact with. Variables defined according to this frame are symbolized using tildes to distinguish them
from those in the Einstein frame which are bare.

Let us consider the case $A(\phi)=e^{\beta \phi^2/2}$ where $\beta$ is a constant.
It is easy to see that $\phi=0$ is a solution of the theory (equivalent to GR),
but surprisingly it is not guaranteed
to be stable in the presence of NSs. Any small perturbation from zero grows and finally leads to
a stable configuration with $\phi \neq 0$. This is called spontaneous scalarization.
The underlying reason for the spontaneous growth becomes clear when we write the linearized
equation of motion (EOM) for small values of $\phi$
\begin{align} \label{scalar_eom}
  \Box_g \phi &= \left( - 8 \pi A^4 \frac{d\left( \ln A(\phi) \right)}{d(\phi^2)} \tilde{T} + m^2_\phi \right)\phi \nonumber\\
  &\approx  \left( - 4 \pi \beta \tilde{T} + m^2_\phi \right)\phi = -\mu_\phi^2 \phi
\end{align}
where $\tilde{T}$ is the trace of the matter stress-energy tensor in the Jordan frame.
For a perfect fluid, $\tilde{T}=-\tilde{\rho}+3\tilde{p}$,
where $\tilde{\rho}$ and $\tilde{p}$ are the rest-frame density and pressure
of the fluid respectively. If the fluid is not strongly relativistic, $\tilde{\rho}\gg\tilde{p}$ and
$\tilde{T}\approx-\tilde{\rho}<0$. The choice $\beta<0$ leads to real $\mu_\phi$ for small enough
$\mphi$, which makes Eq.~\ref{scalar_eom} a massive scalar with the 
``wrong'' sign for the mass square term: a tachyon. This means all Fourier modes of $\phi$ with
wavelength $\lambda \gtrsim1/ \mu_\phi$ and which also fit within the region where the
EOM is tachyonic grow exponentially in amplitude rather than oscillate. This is eventually regulated by the nonlinear 
terms  we ignored, and lead to a finite stable scalar field configuration. Even though we explained
the physics using the linearized equations fully non-linear calculations show that
this scenario is realized for compact stars~\cite{PhysRevLett.70.2220,Ramazanoglu:2016kul}.

Even though we used a specific form of $A(\phi)$, any function with similar next-to-leading quadratic dependence
on $\phi$ gives qualitatively the same results, hence spontaneous growth is generic
for a wide class of nonminimal couplings~\cite{PhysRevLett.70.2220}. We should add that the original calculation used $\mphi=0$.
Indeed, this term actually inhibits spontaneous growth.
However, a nonzero term is needed for agreement with recent binary star
observations~\cite{2013Sci...340..448A, Ramazanoglu:2016kul}.

In short, spontaneous growth depends on two main conditions:
\begin{itemize}
\item R.H.S of Eq.~\ref{scalar_eom} should have a negative coefficient 
in some part of spacetime for a tachyonic degree of freedom,
\item The part of spacetime where the EOM is tachyonic should be
large enough to contain a wavelength of the tachyon $\lambda \sim 1/\mu_\phi$
\end{itemize}
These conditions are
realized for NSs for  order-of-unity values of $|\beta|$. Order of magnitude
calculations show that NSs scalarize for $\beta \lesssim -3$, and scalarization
strengthens as $\beta$ becomes more negative. For $m_\phi$,
the lowest limit is imposed by having no observable effect in relatively close binary systems~\cite{2013Sci...340..448A}
and the upper limit is due to the constraint that the EOM for the scalar, Eq.~\ref{scalar_eom},
should be tachyonic. Overall, this gives the bounds $10^{-9} {\rm eV} \gtrsim m_\phi \gg 10^{-16} {\rm eV}$.

To summarize, the spontaneous growth of a scalar in a scalar-tensor theory depends on a tachyonic EOM
which is a result of the conformal scaling having the expansion 
$A(\phi) \approx 1 -|\beta| \phi^2 + \ldots$.\footnote{Relativistic
matter can satisfy $\tilde{T}>0$ in which case $\beta>0$ can also lead to a tachyonic equation~\cite{Mendes:2014ufa}. 
We will only discuss the $\beta<0$ case, but all our results can be easily adapted to this 
part of the parameter space.} This observation is the key to generalize
spontaneous growth to other nonminimally coupled tensors. Namely, if we replace the scalar
with another tensor field, an ``inverse parabola'' dependence on the tensor field in the
conformal factor leads to a tachyonic EOM where $\tilde{T}$ is not zero.
\section{Spontaneous Vectorization}\label{sec_sv}
Let us replace the scalar in  Eq.~\ref{st_action} with a massive vector $X_\mu$
to obtain a vector-tensor theory action
\begin{align}\label{action_vt}
 \frac{1}{16\pi} &\int d^4x \sqrt{-g} R -  \frac{1}{16\pi} \int d^4x \sqrt{-g} \left[ F^{\mu\nu} F_{\mu\nu}  +2m_X^2 X^\mu X_\mu \right] \nonumber \\
 &+ S_m \left[\psi_m, A_X^2(\eta) g_{\mu \nu} \right], \ \eta =g^{\mu\nu}X_\mu X_\nu
\end{align}
where $F_{\mu\nu} = \nabla_\mu X_\nu -\nabla_\nu X_\mu$,
and the matter terms are analoguous to Eq.~\ref{st_action}. We define the Jordan frame
metric $\tilde{g}_{\mu \nu}=A_X^2 g_{\mu \nu}$, and all variables in this frame are again denoted
with a tilde.
Since spontaneous vectorization has
not been studied before, its observational signatures are not known, including the $m_X=0$ case.
However, our expectation is that $m_X=0$ is not likely to satisfy current observational limits possibly aside from
a marginal part of the parameter space, similar to the case of the massless scalar. Hence,
we consider massive fields. However, we will see in the EOMs that the $m_X$ term antagonizes
tachyonic behavior, i.e. it is actually easier to have a tachyon without it. Thus, our results can be easily
adapted to massless vectors as in the case of the
scalars~\cite{Ramazanoglu:2016kul}. 

The crucial part of the action is again the nonminimal
coupling due to the  conformal scaling $A_X^2(\eta)$, in direct analogy to spontaneous
scalarization. EOMs are
\begin{align} \label{eom_vt}
R_{\mu\nu} &=8\pi \left(T_{\mu\nu} -\frac{1}{2} Tg_{\mu\nu}  \right)  \nonumber \\
 +2 &F_{\mu\rho} F_\nu^\rho -\frac{1}{2} F_{\rho\sigma} F^{\rho\sigma} g_{\mu\nu}
 +2m_X^2 X_\mu X_\nu \nonumber \\
\nabla_\rho F^{\rho \mu}& = \left(-8\pi A_X^4 \Lambda \tilde{T} +m_X^2 \right) X^\mu 
\end{align}
where $\Lambda(\eta) = d (\ln A_X(\eta))/d(\eta)$ and $T_{\mu \nu}$ is the stress-energy tensor in the Einstein frame 
which is related to its counterpart in the Jordan frame through
\begin{equation} \label{Tin_frames}
T_{\mu\nu} \equiv \frac{-2}{\sqrt{-g}}\frac{\delta S_M}{\delta g^{\mu\nu}} = A_X^2 \tilde{T}_{\mu \nu} - 2\Lambda A_X^4 \tilde{T} X_\mu X_\nu\ .
\end{equation}
Also note that Eq.~\ref{eom_vt} implies the constraint
\begin{equation} \label{lorenz_vt}
\left(-8\pi A_X^4 \Lambda \tilde{T} +m_X^2 \right) \nabla_\mu X^\mu = X^\mu \nabla_\mu (-8\pi A_X^4 \Lambda \tilde{T})\  .
\end{equation}

EOM for $X$ is that of a massive vector with negative mass square when
\begin{equation} \label{lorenz_vt}
8\pi A_X^4 \Lambda \tilde{T} > m_X^2\ .
\end{equation}
Hence, we expect the $X=0$ solution to be unstable
and vector fields to spontaneously grow around NSs similar to the case of the scalar.
We will restrict ourselves to $A_X(\eta) = e^{\beta_X \eta/2}$ with constant $\beta_X$ in the following analysis,
but we expect any function with a similar behavior to lead to spontaneous growth as explained in Sec.~\ref{sec_st}.
With this restriction all possible vector-tensor theories
are specified by the two parameters $\beta_X$ and $m_X$. 
For the typical case of $\tilde{T}<0$ this means $\beta_X<0$ of large enough absolute values gives
rise to a tachyon, signalling spontaneous growth of $X$. 

Eq.~\ref{action_vt} has been studied in the literature in a cosmological context,
emphasizing the screening effect of the $m_X$ term~\cite{BeltranJimenez:2013fca}. 
We are mainly concerned with the study of compact objects and strong-field effects in this paper. 
Mass of the vector field is a secondary concern for us compared to the existence of spontaneous 
growth due to a tachyonic instability.
Moreover, we should re-emphasize that the central idea of the paper is the generalization of spontaneous scalarization to 
other tensors, of which the vector case is a specific example.

As  a concrete example, we examine the EOM
for the spacetime of a static, spherically symmetric star with the metric ansatz
\begin{equation}\label{metric_ansatz}
g_{\mu\nu} dx^{\mu} dx^{\nu} = -e^{\nu(r)} dt^2 + \frac{dr^2}{1-2\mu(r)/r} + r^2 d\Omega^2 \ ,
\end{equation}
and the fluid stress energy tensor
\begin{equation}\label{fluid_eq}
\tilde{T}^{\mu\nu}=(\rt+\pt)\tilde{u}^{\mu}\tilde{u}^{\nu}+\pt \tilde{g}^{\mu\nu}\ \ , \ \ \tilde{\nabla}_\mu T^{\mu\nu}=0 
\end{equation}
where the energy density $\rt$, pressure $\pt$, and components of the fluid 4-velocity $\tilde{u}^{\alpha}$
only depend on the radial coordinate $r$. In the absence of any matter, spontaneous vectorization is
identical to a Proca field in general relativity, where the only non-vanishing component of the vector
field for a static, spherically symmetric spacetime is $X_0$~\cite{Herdeiro:2016tmi,Gottlieb:1984jg}.
We will consider the excitation of this component alone, which bears many mathematical similarities
to spontaneous scalarization~\cite{Ramazanoglu:2016kul}. However other components, e.g. $X^r$, might also
be present in the most general case since the potential term for $X$ is not exactly that of a Proca field. 
It is convenient to define new variables $\Phi=-n^\mu X_\mu =-e^{-\nu/2}X_0$ and
$E=(\delta^\mu_{\ r} +n^\mu n_r)F_{\mu\nu} n^{\nu}=e^{-\nu/2}\partial_r X_0$, where
$n_\mu=(-e^{\nu/2},0,0,0)$ is the normal vector to the spatial hypersurfaces. 
Under these assumptions, Eq.~\ref{eom_vt} and~\ref{fluid_eq} reduce to
a modified version of the Tolman-Oppenheimer-Volkoff equations:
\begin{align} \label{tov_vt2}
 \mu' &=  r^2\bigg[ 4\pi A_X^4 \tilde{\rho} -8\pi A_X^4 \Lambda (-\tilde{\rho}+3\tilde{p})\Phi^2  \nonumber\\
 & +\frac{1}{2}(1-\frac{2\mu}{r}) E^2 +\frac{1}{2}m_X^2 \Phi^2 \bigg] \nonumber \\
 \nu' &=\frac{ r^2}{r-2\mu}\left[ 8\pi A_X^4 \tilde{p}-(1-\frac{2\mu}{r}) E^2 +m_X^2 \Phi^2
  +\frac{2\mu}{r^3} \right]   \nonumber \\
\Phi'&= -\frac{\nu'}{2} \Phi -E \nonumber \\
 E' &=\frac{1}{r-2\mu} \big\{ \left[ -2+3\mu/r+\mu' \right] E  \nonumber \\
 & - \left[ m_X^2 -8\pi A_X^4 \Lambda(-\rt+3\pt) \right] r \Phi \big\} \nonumber \\ 
  \pt' &= -(\rt+\pt)\left[ \nu'/2- 2\Lambda \Phi \Phi' \right]
\end{align}
where $'$ is the $r$ derivative. The system is closed by the equation of state (EOS) of
the NS matter $\tilde{\rho}(\tilde{p})$.
\section{Spontaneously Vectorized Neutron Stars}\label{sv_results}
In this section we discuss the 
properties of vectorized NSs for various points on the $\beta_X-m_X$ parameter space.
All past studies on spontaneous scalarization in principle can be  repeated for vectorization, leading to 
many research paths. Here, we limit our investigation to only the most basic aspects of 
spontaneous vectorization.

Eq.~\ref{tov_vt2} can be numerically solved to find isolated, non-rotating, vectorized NSs
using basic numerical techniques detailed in \cite{Ramazanoglu:2016kul}. 
We expect these solutions to be the end states of spontaneously growing vector fields.
Vectorization depends on the equation of state of the NS matter, and we detail the results for 
the intermediately stiff ``HB EOS'' as defined in \cite{2009PhRvD..79l4033R}. 
Vectorization  qualitatively behaves the same way for stiffer and softer equations in
\cite{2009PhRvD..79l4033R} as well. We use $A_X(\Phi(r=0))-1\equiv A_0-1$, as a measure of
the strength of vectorization, which also quantifies the magnitude of deviations from GR.
The solutions we present are tested by the convergence of the Hamiltonian
constraint to zero as the grid resolution is increased in a three dimensional initial
data solver~\cite{idsolve_paper}.

\begin{figure}
\includegraphics[width=8.5cm,left]{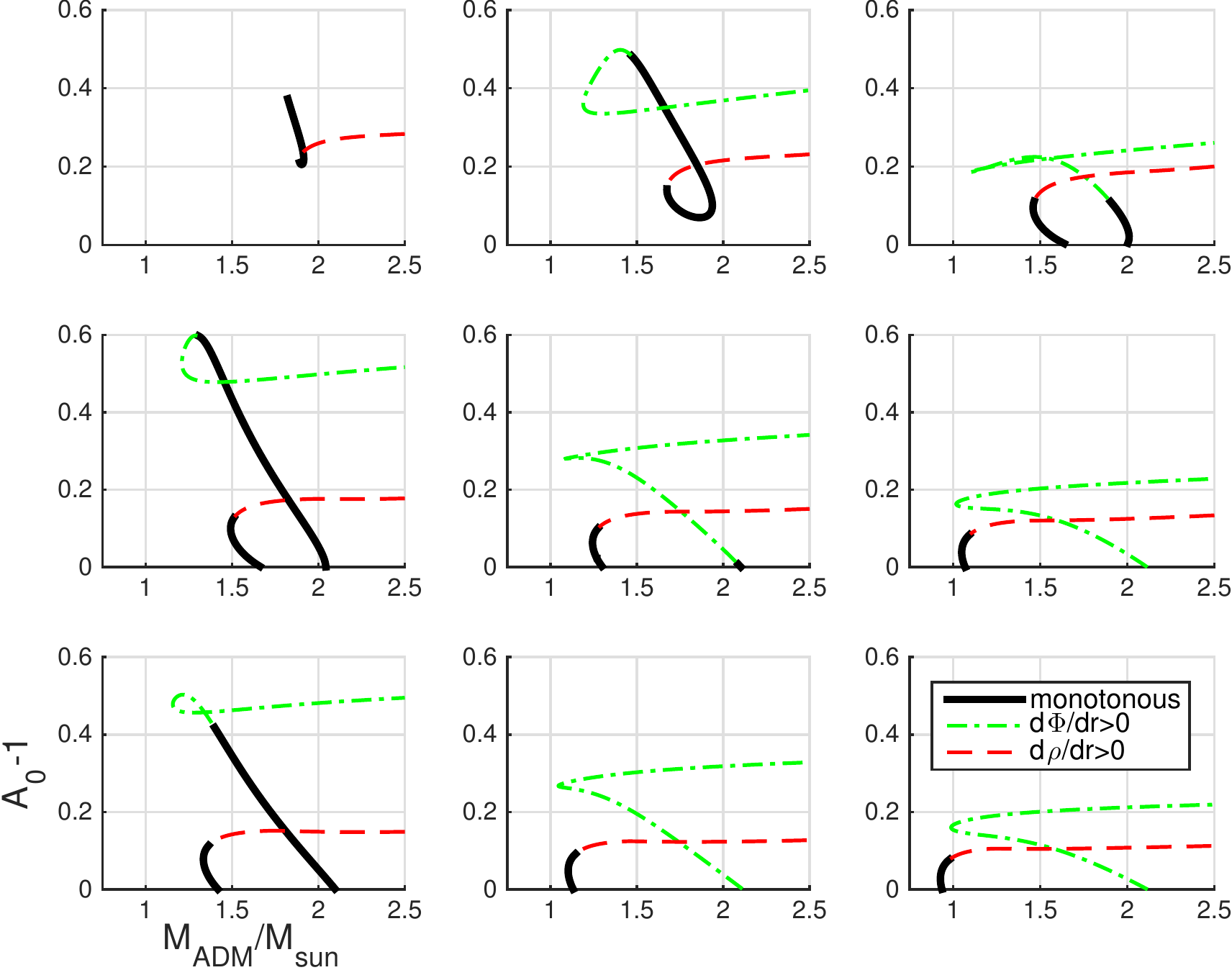}
\caption{
Effect of varying $m_X$ and $\beta_X$ on the strength of vectorization, i.e.
$A_0-1$ vs $M/M_\odot$ plots.
Upper row: $m_{X}= 1.6 \times 10^{-11}  \textrm{eV}$, middle 
row: $m_{X}= 8.0 \times 10^{-12}$, lower row: $m_{X}= 4.8 \times 10^{-12} \textrm{eV}$, 
left column: $\beta_X =-4$, middle column: $\beta_X=-5$, right column: $\beta_X=-6$. 
Dashed and dot-dashed lines indicate solution where $\tilde{\rho}$ or $\Phi$ does not
monotonically decrease with radius, respectively. 
We also suspect some of the monotonically decreasing solutions 
(solid lines) to be unstable.
}
\label{parameter_space}.
\end{figure}

Vectorization of NSs as a function of their Arnowitt-Deser-Misner (ADM) mass, $M_{ADM}$, 
for various values of $\beta_X$ and $m_X$ can be seen in Fig.~\ref{parameter_space}.
Vectorization does not occur for $\beta_X \gtrsim -2$, and gets weaker with increasing
$|\beta_X|$ for $\beta_X<-4$, whereas spontaneous scalarization gets stronger for highly negative
$\beta$ values~\cite{Ramazanoglu:2016kul}. This is possibly related to the fact that $A_X>1$ for 
vectorization ($\eta=X_\mu X^\mu=-\Phi^2 <0$) while $A<1$ for scalarization, but this point
calls for further investigation.

$M_{ADM}-(A_0-1)$ graphs become qualitatively different as $m_X$ changes for a given $\beta_X$, but there is not 
a clear relationship in terms of the strength of the vector field. This is also different from
spontaneous scalarization where the scalar field monotonously becomes weaker with increasing
$\mphi$~\cite{Ramazanoglu:2016kul}. Overall, relationship between $M_{ADM}$ and the spontaneously 
growing field seems to be more complex
for vectorization compared to scalarization.
 
We believe at least part of this complexity in Fig.~\ref{parameter_space}. to be misleading. Even though all points in
the figure represent solutions to Eq.~\ref{tov_vt2}, their stability to small
perturbations are not guaranteed. We suspect many of the solutions to be unstable based on
our experience with spontaneously scalarized stars~\cite{Ramazanoglu:2016kul}.
For some stars $\Phi$ or $\rho$ are not
monotonously decreasing with radius (non-solid lines in Fig.~\ref{parameter_space}.), 
and in general we expect only a single solution to be stable for a given $M_{ADM}$ value.
If these expectations are true, then it is likely that vectorization monotonically gets weaker with
increasing $m_X$. However, a complete answer to the stability question of the solutions we
found require numerical time evolution of these stars which is left for future work.

Allowed field mass values are estimated similarly for spontaneous scalarization
and vectorization.
The vector field dies off exponentially away from the star with a decay length of $1/m_X$.
Thus the lower bound to $m_X$ comes from the lack of observational signatures of modifications to GR 
in binary systems, $m_X \gg 10^{-16} {\rm eV}$~\cite{Ramazanoglu:2016kul,2013Sci...340..448A}. The upper
bound criterion for $m_X$ is having a tachyon in the EOM of the vector field in Eq.~\ref{eom_vt}, which translates to
$10^{-9} {\rm eV} \gtrsim m_X$. We should note that spontaneous vectorization seems to continue to be effective
at somewhat higher field masses compared to spontaneous scalarization, but an exact verdict depends on the
stability of the vectorized NSs at these high $m_X$ values.

\begin{figure}
\includegraphics[width=8.5cm,left]{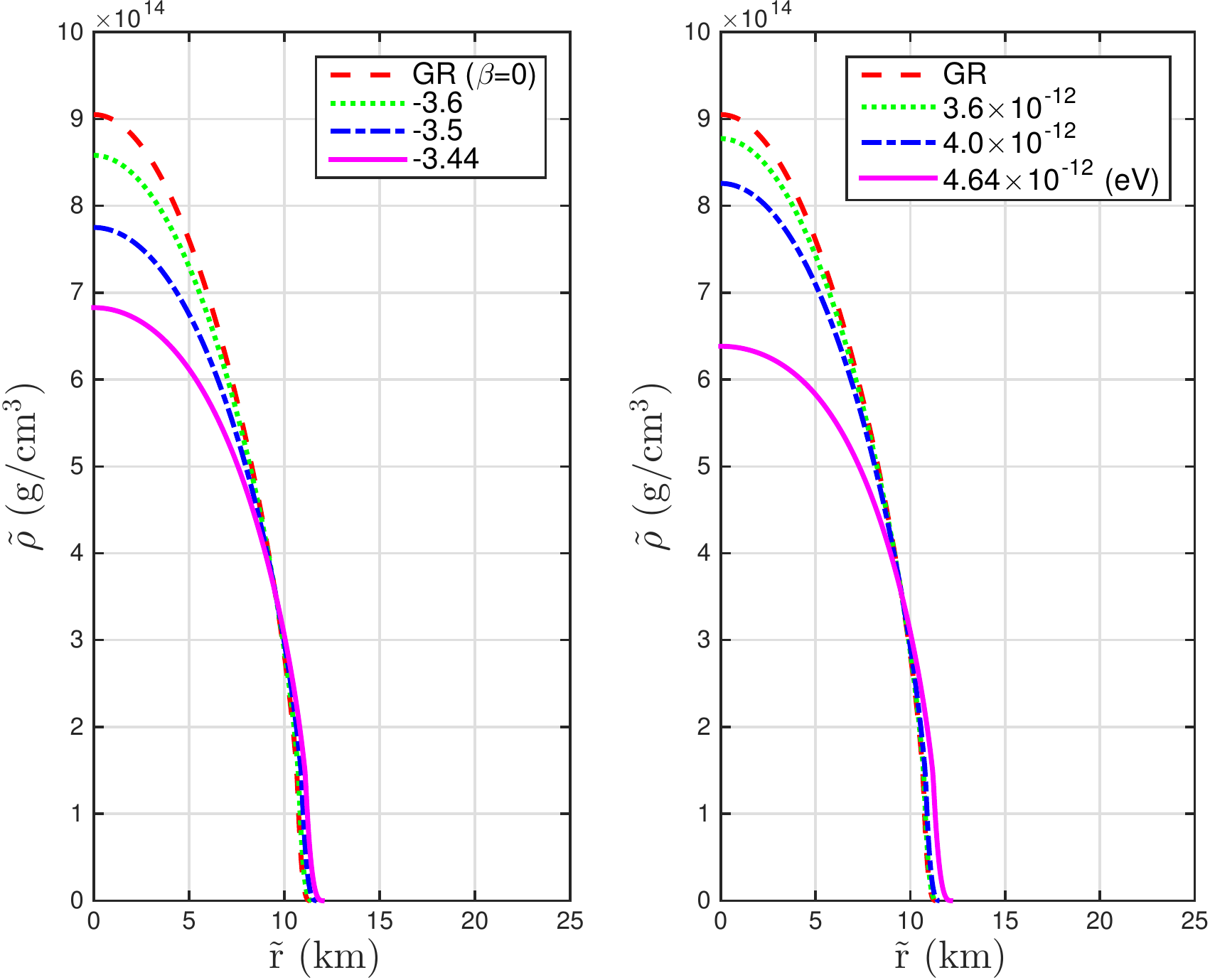}
\caption{
Changes in NS structure with spontaneous vectorization for various stars with 
$M_{ADM}=1.35 M_\odot$.
Left: Effect of varying $\beta_X$  for $m_X=1.6 \times 10^{-13}  \textrm{eV}$.
Right: Effect of varying $m_X$  for $\beta_X=-4$
}
\label{star_profile}.
\end{figure}

Density profiles of various vectorized stars we expect to be stable can be seen in Fig.~\ref{star_profile}. 
There is significant deviation from the predictions of GR, but unlike spontaneous scalarization the central density is less than
what one would have in GR. This is again possibly due to the fact that $A_X>1$ for vectorization. Despite significant
changes in the density profiles, the radii or baryon masses of the stars did not change appreciably from the values in
GR, in contrast to spontaneous scalarization~\cite{Ramazanoglu:2016kul}.

Signs of spontaneous vectorization are easily distinguishable in astrophysical observables thanks to the non-perturbative nature of
the deviations from GR. We expect compact object binaries containing a vectorized star to have significantly 
different evolution near the merger. Another possible observable signature of spontaneous vectorization is
dynamical vectorization, where a weakly vectorized NS significantly increases its vector field
upon the influence of an approaching strongly vectorized NS~\cite{PhysRevD.89.044024}. 
Observations of isolated NSs can also be useful in constraining spontaneous vectorization 
parameter space~\cite{Palenzuela:2015ima}.
Lastly, spontaneous scalarization
has also been recently investigated for collapsing stars, which is another avenue of research
for vectorization~\cite{Gerosa:2016fri}.
\section{Conclusions and Future Work}\label{sv_conclusions}
In this study, we show that spontaneous scalarization is a specific case of a 
more general phenomenon that arises in gravity theories with nonminimal coupling to matter. 
The key fact in generalizing spontaneous scalarization is that the growth of the scalar field 
is due to the form of the nonminimal coupling terms and not due to the nature of the field itself. Other fields with
similar nonminimal coupling terms can easily be shown to have the same tachyonic behavior as the
spontaneously growing scalars which leads to growth.
Hence, classifying spontaneous scalarization as a specific case of spontaneous tensorization
illuminates the underlying physics better than describing it as an unexpected phenomena in certain scalar-tensor theories.

The most straightforward generalization of spontaneous scalarization is replacing the scalar with a vector: spontaneous vectorization.
We derived the EOMs for such a theory, and used them to construct spontaneously vectorized NSs. 
We showed that spontaneous vectorization can lead to 
order-of-unity changes in NS structure which strongly indicates prominent observable signatures in compact object mergers, and
possibly isolated NSs.
Consequently, observations of strong-field gravity can confirm or rule out this nonminimally coupled theory with relative ease. 
Spontaneous vectorization of isolated spherically symmetric NSs is somewhat
more complex than the case of spontaneous scalarization, and a complete understanding of their stability will
require numerical time evolution~\cite{Zilhao:2015tya}.

We can devise a ``recipe'' for a theory of spontaneous growth of any given field $\xi$ with an action similar to Eq.~\ref{st_action}. 
The conformal factor should be schematically in the form $A_\xi(\xi)=1+\beta_\xi \xi^2+ \dots$, in which case $\beta_\xi$ with the ``correct'' sign
leads to a tachyonic EOM, hence spontaneous growth. For example, spontaneous
growth readily generalizes to complex scalars and vectors. There is also an example of this mechanism
for a $3$-form field  which utilizes the very close relationship between
$3$-forms and vectors in $4$ dimensional spacetimes~\cite{Barreiro:2016aln}.

The recipe we provide may require special care for the next case in the tensor hierarchy,
a rank-2 tensor. 
Gravity theories with two interacting spin-2 fields, especially when the fields are massive, have been notoriously
hard to approach~\cite{deRham:2014zqa}. For example,
choosing a scalar analog of $\eta=X_\mu X^\mu$ for rank-2 tensors
which can act as the argument of $A_\xi$ is a subtle issue~\cite{deRham:2014zqa}. 
Despite historical problems, there has been recent
breakthroughs in theories of massive rank-2 tensors in terms of massive gravity and bigravity,
and we are now able to build anomaly-free theories of interacting rank-2 tensors~\cite{deRham:2014zqa}. 
We aim to use such theories as starting points to formulate spontaneous growth for rank-2 tensors while
taking care of  known issues in extending massive gravity theories~\cite{deRham:2014fha}. Spontaneous growth of spinor
fields is another avenue to investigate.

Spontaneous scalarization has been a viable theory of gravity for more than two decades, and its promise of 
non-perturbative deviations from GR at the strong-field regime has made it even more appealing at the age
of gravitational wave observations. We expect generic spontaneous tensorization theories to share these important 
traits with scalarization, and open new possibilities in strong gravity research.

\acknowledgments
This project was supported by the Scientific and Technological Research Council of Turkey (T\"{U}B\.{I}TAK) 
Co-funded Brain Circulation Scheme (Co-Fund) project number 116C033. Numerical calculations were performed on the Perseus
cluster at Princeton University, COSMOS cluster at the University of Cambridge, and L\"{u}fer cluster at Ko\c{c} University.
Part of this work was undertaken while the author was a postdoctoral researcher at DAMTP, University of Cambridge, where the
computational work was supported by the Cosmos Shared Memory system through BIS Grant
No.~ST/J005673/1 and STFC Grant Nos.~ST/H008586/1, ST/K00333X/1.
We thank William East and Ulrich Sperhake for valuable discussions.

\bibliographystyle{h-physrev}
\bibliography{st}

\begin{thebibliography}{10}

\bibitem{PhysRevLett.70.2220}
T.~Damour and G.~Esposito-Far\`ese,
\newblock Phys. Rev. Lett. {\bf 70}, 2220 (1993).

\bibitem{Will:2006LR}
C.~M. Will,
\newblock Living Rev. Relativity {\bf 9} (2006), {arXiv:gr-qc/0510072},
\newblock \url{http://www.livingreviews.org/lrr-2006-3}.

\bibitem{2013PhRvD..87h1506B}
E.~{Barausse}, C.~{Palenzuela}, M.~{Ponce}, and L.~{Lehner},
\newblock \prd {\bf 87}, 081506 (2013), 1212.5053.

\bibitem{2014PhRvD..89h4005S}
M.~{Shibata}, K.~{Taniguchi}, H.~{Okawa}, and A.~{Buonanno},
\newblock \prd {\bf 89}, 084005 (2014), 1310.0627.

\bibitem{Ramazanoglu:2016kul}
F.~M. Ramazano\u{g}lu and F.~Pretorius,
\newblock Phys. Rev. {\bf D93}, 064005 (2016), 1601.07475.

\bibitem{Horava:2009uw}
P.~Horava,
\newblock Phys. Rev. {\bf D79}, 084008 (2009), 0901.3775.

\bibitem{Emparan2008}
R.~Emparan and H.~S. Reall,
\newblock Living Reviews in Relativity {\bf 11}, 6 (2008).

\bibitem{Jacobson:2000xp}
T.~Jacobson and D.~Mattingly,
\newblock Phys. Rev. {\bf D64}, 024028 (2001), gr-qc/0007031.

\bibitem{Moffat:2005si}
J.~W. Moffat,
\newblock JCAP {\bf 0603}, 004 (2006), gr-qc/0506021.

\bibitem{Bekenstein:2004ne}
J.~D. Bekenstein,
\newblock Phys. Rev. {\bf D70}, 083509 (2004), astro-ph/0403694,
\newblock [Erratum: Phys. Rev.D71,069901(2005)].

\bibitem{Chamseddine:2013kea}
A.~H. Chamseddine and V.~Mukhanov,
\newblock JHEP {\bf 11}, 135 (2013), 1308.5410.

\bibitem{Sebastiani:2016ras}
L.~Sebastiani, S.~Vagnozzi, and R.~Myrzakulov,
\newblock Adv. High Energy Phys. {\bf 2017}, 3156915 (2017), 1612.08661.

\bibitem{BeltranJimenez:2013fca}
J.~Beltran~Jimenez, A.~L. Delvas~Froes, and D.~F. Mota,
\newblock Phys. Lett. {\bf B725}, 212 (2013), 1212.1923.

\bibitem{Arvanitaki:2009fg}
A.~Arvanitaki, S.~Dimopoulos, S.~Dubovsky, N.~Kaloper, and J.~March-Russell,
\newblock Phys. Rev. {\bf D81}, 123530 (2010), 0905.4720.

\bibitem{Kodama:2011zc}
H.~Kodama and H.~Yoshino,
\newblock Int. J. Mod. Phys. Conf. Ser. {\bf 7}, 84 (2012), 1108.1365.

\bibitem{Jaeckel:2010ni}
J.~Jaeckel and A.~Ringwald,
\newblock Ann. Rev. Nucl. Part. Sci. {\bf 60}, 405 (2010), 1002.0329.

\bibitem{Goodsell:2009xc}
M.~Goodsell, J.~Jaeckel, J.~Redondo, and A.~Ringwald,
\newblock JHEP {\bf 11}, 027 (2009), 0909.0515.

\bibitem{PhysRevLett.116.061102}
LIGO Scientific Collaboration and Virgo Collaboration, B.~P. Abbott {\em
  et~al.},
\newblock Phys. Rev. Lett. {\bf 116}, 061102 (2016).

\bibitem{2013Sci...340..448A}
J.~{Antoniadis} {\em et~al.},
\newblock Science {\bf 340}, 448 (2013), 1304.6875.

\bibitem{Mendes:2014ufa}
R.~F.~P. Mendes,
\newblock Phys. Rev. {\bf D91}, 064024 (2015), 1412.6789.

\bibitem{Herdeiro:2016tmi}
C.~Herdeiro, E.~Radu, and H.~Runarsson,
\newblock Class. Quant. Grav. {\bf 33}, 154001 (2016), 1603.02687.

\bibitem{Gottlieb:1984jg}
D.~Gottlieb, R.~Hojman, L.~H. Rodriguez, and N.~Zamorano,
\newblock Nuovo Cim. {\bf B80}, 62 (1984).

\bibitem{2009PhRvD..79l4033R}
J.~S. {Read} {\em et~al.},
\newblock \prd {\bf 79}, 124033 (2009), 0901.3258.

\bibitem{idsolve_paper}
W.~E. East, F.~M. Ramazano\u{g}lu, and F.~Pretorius,
\newblock Phys. Rev. {\bf D86}, 104053 (2012), 1208.3473.

\bibitem{PhysRevD.89.044024}
C.~Palenzuela, E.~Barausse, M.~Ponce, and L.~Lehner,
\newblock Phys. Rev. D {\bf 89}, 044024 (2014).

\bibitem{Palenzuela:2015ima}
C.~Palenzuela and S.~Liebling,
\newblock Phys. Rev. {\bf D93}, 044009 (2016), 1510.03471.

\bibitem{Gerosa:2016fri}
D.~Gerosa, U.~Sperhake, and C.~D. Ott,
\newblock Class. Quant. Grav. {\bf 33}, 135002 (2016), 1602.06952.

\bibitem{Zilhao:2015tya}
M.~Zilhão, H.~Witek, and V.~Cardoso,
\newblock Class. Quant. Grav. {\bf 32}, 234003 (2015), 1505.00797.

\bibitem{Barreiro:2016aln}
T.~Barreiro, U.~Bertello, and N.~J. Nunes,
\newblock (2016), 1610.00357.

\bibitem{deRham:2014zqa}
C.~de~Rham,
\newblock Living Rev. Rel. {\bf 17}, 7 (2014), 1401.4173.

\bibitem{deRham:2014fha}
C.~de~Rham, L.~Heisenberg, and R.~H. Ribeiro,
\newblock Phys. Rev. {\bf D90}, 124042 (2014), 1409.3834.

\end{thebibliography}

\end{document}